\documentclass[10pt,twocolumn,letterpaper]{article}

\usepackage[pagenumbers]{cvpr} 

\usepackage[dvipsnames]{xcolor}

\definecolor{myblue}{rgb}{0.21,0.49,0.74}
\usepackage[pagebackref,breaklinks,colorlinks,citecolor=myblue]{hyperref}

\usepackage{graphicx,subcaption,caption}
\usepackage{multirow}
\usepackage{amssymb}
\usepackage{makecell}
\usepackage{utfsym}
\usepackage{pifont}
\usepackage{colortbl}
\usepackage{color, xcolor} 
\usepackage{tcolorbox}
\usepackage{listings}

\definecolor{darkred}{rgb}{0.6,0.0,0.0}
\definecolor{darkgreen}{rgb}{0,0.50,0}
\definecolor{lightblue}{rgb}{0.0,0.42,0.91}
\definecolor{orange}{rgb}{0.99,0.48,0.13}
\definecolor{grass}{rgb}{0.18,0.80,0.18}
\definecolor{pink}{rgb}{0.97,0.15,0.45}

\definecolor{codegreen}{rgb}{0,0.6,0}
\definecolor{codegray}{rgb}{0.5,0.5,0.5}
\definecolor{codepurple}{rgb}{0.58,0,0.82}
\definecolor{backcolour}{rgb}{0.95,0.95,0.92}

\lstdefinestyle{mystyle}{
  frame=single,
  basicstyle=\ttfamily\footnotesize,
  backgroundcolor=\color{backcolour}, commentstyle=\color{codegreen},
  commentstyle=\color{darkgreen}\slshape,
  keywordstyle=\color{blue},
  stringstyle=\color{darkred},
  numberstyle=\tiny\color{codegray},
  emphstyle=\color{pink}\underbar,
  morekeywords={Verify, Question},
  escapeinside={(*@}{@*)},
  breakatwhitespace=false,         
  breaklines=true,                 
  captionpos=b,                    
  keepspaces=true,                    
  numbersep=5pt,                  
  showspaces=false,                
  showstringspaces=false,
  showtabs=false,                  
  tabsize=2
}

\usepackage{pifont}
\newcommand{\cmark}{\ding{51}}%
\newcommand{\xmark}{\ding{55}}%

\newcommand{\snifferemoji}{\includegraphics[height=2.5\fontcharht\font`\B]{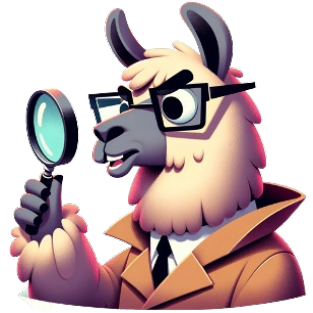}}

\colorlet{dark-green}{green!60!black}

\definecolor{tiffany}{rgb}{.505,.847,.815}

\title{
\snifferemoji{} 
\textsc{Sniffer}: Multimodal Large Language Model \\ for Explainable Out-of-Context Misinformation Detection
\vspace{-.1in}
}

\author{%
{\bf Peng Qi}, 
{\bf Zehong Yan},
{\bf Wynne Hsu}, 
{\bf Mong Li Lee} \\
{National University of Singapore}  \\
{\tt peng.qi@nus.edu.sg, e0787894@u.nus.edu,\{whsu,leeml\}@comp.nus.edu.sg} \\
\textbf{\url{https://pengqi.site/Sniffer}}
\vspace{-.1in}
}

\begin{document}

\twocolumn[{
      \vspace{-1em}
      \maketitle
      \vspace{-1em}
      \begin{center}
        \centering
        \vspace{-0.2in}
        \includegraphics[width=0.9\linewidth]{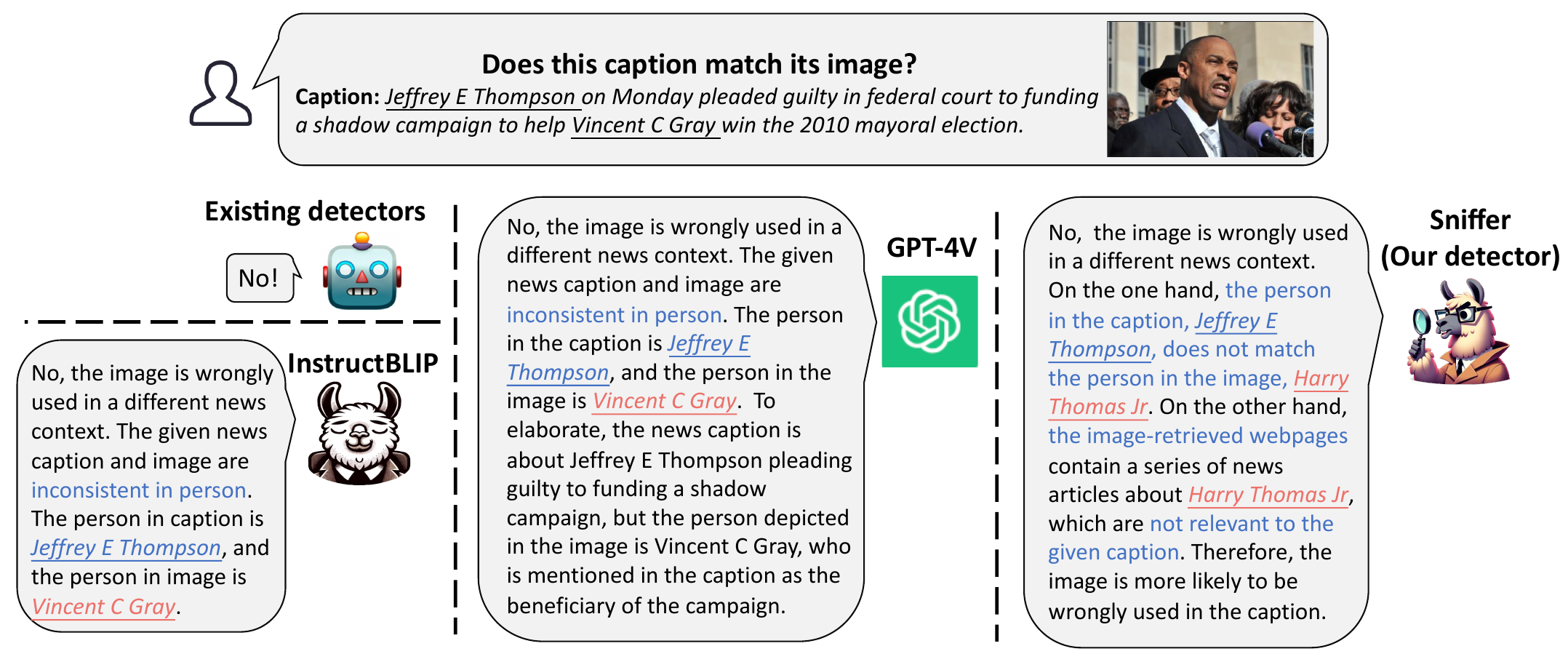}
       \vspace{-.05in}
        \captionof{figure}{
        Comparison between the proposed \textsc{Sniffer} and other detectors\protect\footnotemark. In this out-of-context misinformation, the individual in the image is {\it Harry Thomas Jr}, which contradicts the caption. Existing detectors often give a judgment without explanation. While InstructBLIP and GPT-4V correctly identify the inconsistent news element (\ie person) in the image-text pair, they mistakenly associate the person in the image with a different individual mentioned in the caption. In contrast, \textsc{Sniffer} analyzes both the consistency of the image-text content and the claim-evidence relevance, and accurately identify the person in the image as {\it Harry Thomas Jr}, thereby providing a precise and persuasive explanation.
        }
        \label{fig:introcase}
      \end{center}
    }]

   \footnotetext{Due to space constraints, we have made minor edits to the models' responses for brevity without altering their original meaning.}
    
\begin{abstract}
Misinformation is a prevalent societal issue due to its potential high risks. Out-Of-Context (OOC) misinformation, where authentic images are repurposed with false text, is one of the easiest and most effective ways to mislead audiences.
Current methods focus on assessing image-text consistency but lack convincing explanations for their judgments, which is essential for debunking misinformation. 
While Multimodal Large Language Models (MLLMs)  have rich knowledge and innate capability for visual reasoning and explanation generation,
they still lack sophistication in understanding and discovering the subtle cross-modal differences. 
In this paper, we introduce \textsc{Sniffer}, a novel multimodal large language model specifically engineered for OOC misinformation detection and explanation. 
\textsc{Sniffer}  employs two-stage instruction tuning on InstructBLIP. 
The first stage refines the model's concept alignment of generic objects with news-domain entities and the second stage leverages language-only GPT-4 generated OOC-specific instruction data to fine-tune the model's discriminatory powers.
Enhanced by external tools and retrieval, \textsc{Sniffer} not only detects inconsistencies between text and image but also utilizes external knowledge for contextual verification. 
Our experiments show that \textsc{Sniffer} surpasses the original MLLM by over 40\% and outperforms state-of-the-art methods in detection accuracy. \textsc{Sniffer} also provides accurate and persuasive explanations as validated by quantitative and human evaluations. 

\end{abstract}    
\vspace{-.1in}
\section{Introduction}
\label{sec:intro}

In recent years, Deepfake and other media manipulation technologies \cite{dgm, deepfake21, deepfake22, deepfakesurvey} have garnered considerable attention from both the computer vision community and the general public, due to their lifelike qualities and their significant influence in accelerating the spread of fake news.
Nevertheless, one of the easiest and most prevalent ways to mislead audiences is to use unaltered images in a new but false or misleading context, known as out-of-context (OOC) misinformation \cite{bg-ooc}. 
For example, during the recent Israel-Hamas war, numerous instances of OOC misinformation were observed on social media. These often involved repurposing old images from unrelated armed conflicts or even military footage from video games \cite{case-ooc}. 
Detecting such OOC misinformation poses a unique challenge since the visual content remains authentic, and the deception stems solely from the context created by combining these images with misleading or incorrect text.

Faced with this challenge, current studies \cite{newsclippings, dttrm, ooc-mm17, ooc-mm18, ooc-cvpr19, ccn, icmr} focus on learning a unified latent representation space to assess the consistency of image-text pairs or compare them with external references. 
While these approaches have made some progress, the development of convincing explanations for these judgments, critical for establishing public trust and more effectively debunking misinformation \cite{bg-explanation, bg-explanation-2}, still remains unexplored.
Although some works \cite{newsclippings, ccn} visualize salient objects and words based on the model's attention weights, it is still unclear why these highlighted regions signify misinformation. 
Therefore, it is important to develop technologies capable of not only detecting but also explaining out-of-context misinformation.

Multimodal large language models (MLLMs) have made great advancements in a variety of multimodal tasks \cite{survey-mllm, survey-mllm-2}. With their extensive world knowledge and robust visual reasoning and generation capabilities, MLLMs have the potential to detect factual inconsistency in image-text pairs and to generate coherent, natural language-based explanations.  
However, {\it applying existing MLLMs to the task of OOC misinformation detection is non-trivial}. 
{\it On the one hand}, initial experiments on some open-sourced MLLMs (such as InstructBLIP \cite{instructblip})
revealed common shortcomings, including failure to follow instructions, misunderstanding of user's intent, and hallucination (detailed in Supplementary).
These shortcomings may stem from the fundamental differences between the training corpus for MLLMs and the requirements of the specific task. In OOC detection, the model must discern when text and images describe disparate news events, which are typically not present in MLLMs' training data. 
{\it On the other hand}, 
the news event from which the image originates may not be discernible from the image itself. For example, news images accompanying reports of public figures speaking at various events are generally similar close-up shots of the individuals, making it almost impossible to distinguish from the image content alone which news scene or event the image is from. 
MLLMs have inherent limitations as they lack the ability to access real-time information or utilize external tools for identifying and contextualizing events.

To address these challenges, we introduce \textsc{Sniffer}, an MLLM specifically engineered for detecting OOC misinformation.
We utilize language-only GPT-4 \cite{gpt4} to generate the instruction data that includes both judgments and explanation, and employ two-stage instruction tuning on InstructBLIP. 
We first refine the model’s concept alignment of generic objects with news-domain entities by image captioning data, and then leverage the OOC-specific data to fine-tune the model’s discriminatory powers. 
Through task-specific tuning, the inherent knowledge within the model can be activated and reorganized to align with the specific logic required by our task. 
This is the first attempt to extend multimodal instruction tuning to the news domain for end-to-end training of an OOC misinformation detector. 

Moreover, \textsc{Sniffer} is augmented with external knowledge through retrieval and tool usage. It conducts {\it internal checking} to spot inconsistencies between text and image, and {\it external checking} to reason between the given text and retrieved image context. The judgments and explanations derived from these two perspectives are then integrated into the LLM module for a more comprehensive and unified output.
Empowered by multimodal instruction tuning and enhanced with external knowledge, \textsc{Sniffer} demonstrates a significant performance improvement, exceeding the original MLLM by over 40\% and surpassing current state-of-the-art methods. Beyond accurate judgments, \textsc{Sniffer} also excels in providing precise and persuasive explanations, as evidenced by quantitative analysis and human evaluations.

Our contribution can be summarized as follows:
\begin{itemize}
    \item We design a novel data reformation pipeline assisted by language-only GPT-4 to convert the given OOC image-text pairs into the appropriate instruction-following format with judgments and explanations simultaneously. 
    \item 
    We propose a practical approach to adapt existing general-purpose MLLMs for out-of-context misinformation detection through two-stage instruction tuning. Our task-specific MLLM, \textsc{Sniffer}, enhanced with external tools and retrieval, effectively models both the internal image-text clues and the external claim-evidence clues for simultaneous OOC detection and explanation.
    \item Extensive experiments show that \textsc{Sniffer} significantly surpasses the original MLLM and SOTA methods in detection performance, achieves comparable results with just 10\% of the training data, and provides precise and persuasive explanations validated by both quantitative and human assessments.
\end{itemize}

\section{Related Work}
\label{sec:related}

{\noindent \bf Out-of-Context Misinformation Detection.}
Existing multimodal misinformation detection methods extract and fuse features from different perspectives, such as linguistic patterns \cite{linguistic}, image tampered patterns \cite{icdm, dgm}, multimodal inconsistency \cite{safe, mm21}, user response \cite{defend}, and propagation structure \cite{maprop}, to classify the given news as real or fake. 
Out-of-context misinformation, also known as image repurposing 
and cheapfake, is a specific form of misinformation that is easy to create and highly misleading.
It reuses authentic images within a similar yet incorrect context, rendering many traditional detection methods ineffective \cite{neu-sym}.

To detect OOC misinformation, some methods utilize knowledge-rich pre-trained models to conduct {\it internal checking} for the given image-text pair. For example, \cite{newsclippings, dttrm} use the multimodal pre-trained models CLIP \cite{clip} and VisualBERT \cite{visualbert} to classify.  
Other methods employ external resources to do {\it external checking}. For example, \cite{ooc-mm17, ooc-mm18, ooc-cvpr19} use a reference dataset that contains unmanipulated related claims to mimic world knowledge, and then detect the OOC use by comparing the given claim with the retrieved one. 
\cite{ccn} uses the text and image to retrieve related Web evidence separately, and  compute the claim-evidence consistency under both textual and visual modalities. 
Similarly, \cite{icmr} proposes an unsupervised cross-modal entity consistency verification method, which retrieves images using entities extracted from the given text, and then calculates the similarity between the entities in the given image and retrieved images.
Different from judging the veracity of the given image-caption claim, \cite{cosmos} re-formulates this task as:
If two captions refer to the same objects in the same image but are semantically different, then it indicates an out-of-context use of this image. They introduce a self-supervised training strategy to train the model's visual grounding ability and evaluate the model on OOC samples. 

Although existing works have made some progress, they often fail to explain their judgment. This lack of transparency hinders the effectiveness of these methods in debunking misinformation.
\cite{neu-sym} proposes an interpretable de-contextualization detector, which uses MLLM to verify the sub-questions decomposed from the given text and selects supported question-answer pairs as the explanation. However, the effectiveness of this method is heavily limited by the ability of existing general-domain MLLM. 
In contrast, we use a two-stage instruction tuning to adapt the general-domain MLLM for the OOC detection task, enabling it to concurrently generate both judgment and explanation. 

\smallskip
{\noindent \bf Instruction Tuning for MLLMs.}
Multimodal large language models typically use a lightweight visual prompt generator that produces soft prompts for the input images to connect to an existing LLM. 
Early models \cite{Flamingo, blip2, minigpt} focus on large-scale pre-training, while recent works \cite{llava, instructblip, otter} employ instruction tuning for pre-trained MLLMs. 
 Instruction tuning is a crucial technique to enhance the capabilities and controllability of large language models, which involves further training pre-trained models on a collection of instruction-formatted datasets to enhance models' generalization to unseen tasks.
 One challenge of instruction tuning is how to construct high-quality instructions that properly cover the desired target behaviors. In addition to adapting existing benchmarks \cite{instructblip}, some works \cite{llava, otter} collect samples through self-instruction \cite{selfinstruct}, which bootstraps LLMs to generate textual instruction-following data using a few hand-annotated samples.
 The success of LLMs in the general domain has inspired interest in applications such as biomedical \cite{llava-med}, law \cite{lawyer} and education \cite{taoli}. In this work, we extend the ability of general-domain MLLM for OOC misinformation detection via instruction tuning. 

\begin{figure*}[bhtp]
	\centering
	\includegraphics[width=\textwidth]{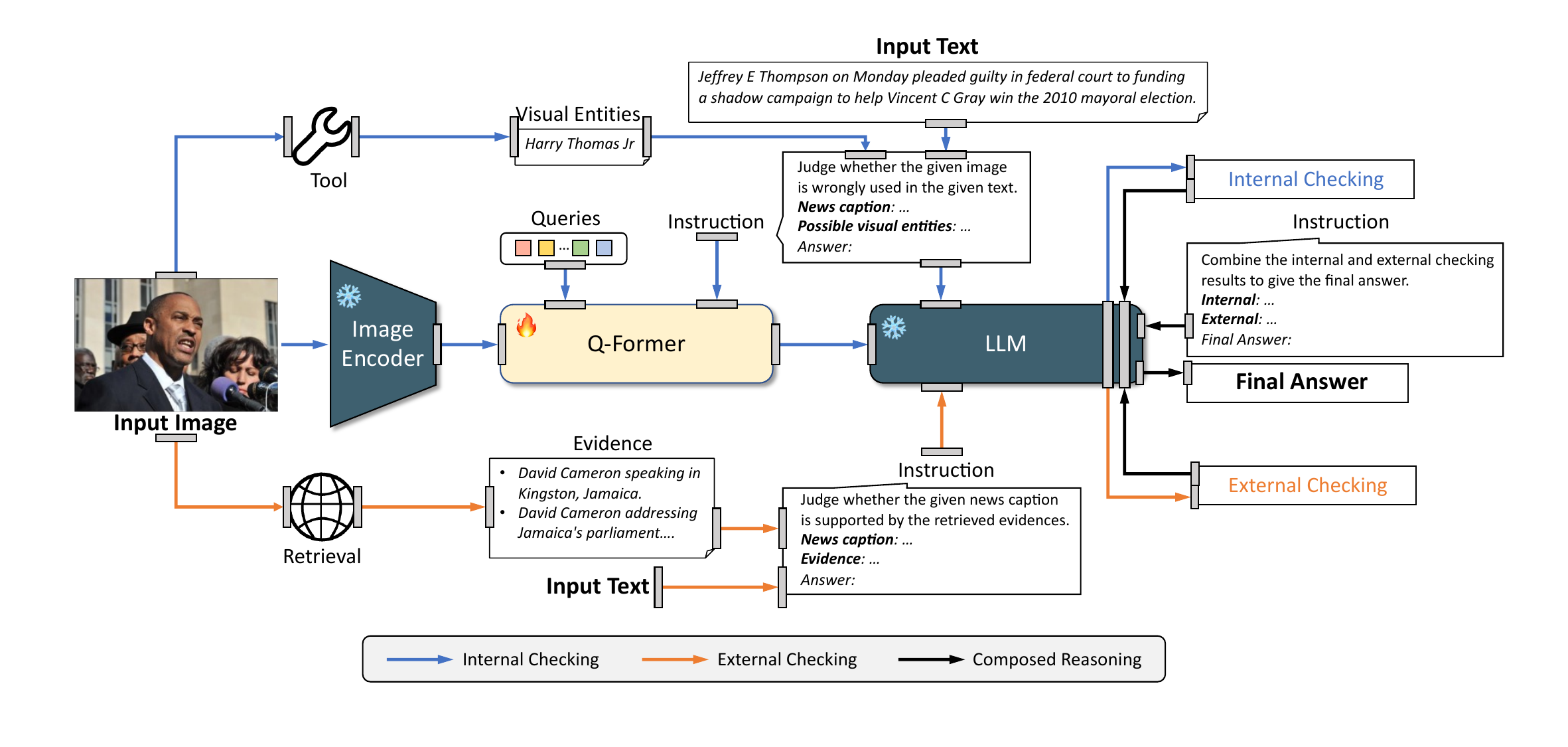}
  \vspace{-.2in}
	\caption{Architecture of the proposed framework \textsc{Sniffer}. For a given image-text pair, \textsc{Sniffer} conducts a two-pronged analysis: (1) it checks the consistency of the image and text content (\textcolor[RGB]{79, 113, 190}{\it internal checking}), and (2) it examines the relevance between the context of the retrieved image and the provided text (\textcolor[RGB]{222, 131, 68}{\it external checking}). The outcomes of both these verification processes are then considered by \textsc{Sniffer} to arrive at a final judgment and explanation.}
	\label{fig:sniffer}
  \vspace{-.1in}
\end{figure*} 

\section{Method}
\label{sec:method}

\begin{figure}[t]
	\centering
	\includegraphics[width=.45\textwidth]{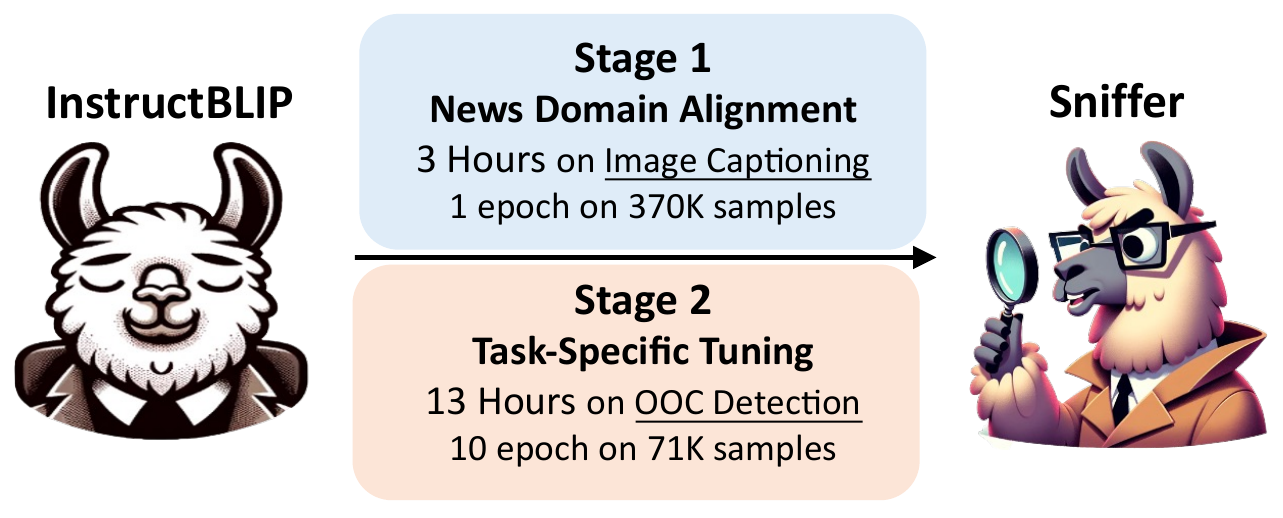}
  \vspace{-.1in}
	\caption{\textsc{Sniffer} was initialized with the general-domain InstructBLIP and then continuously trained to adapt it to the news domain and OOC misinformation detection task sequentially.}
	\label{fig:training}
  \vspace{-.13in}
\end{figure} 

Our goal is to develop an explainable multimodal out-of-context misinformation detection model that can jointly output the prediction and explanation. Figure~\ref{fig:sniffer} illustrates the overall architecture of the proposed \textsc{Sniffer}. Specifically, for a given image-text pair, we feed it into the multimodal large language model for checking the internal cross-modal inconsistencies. The image-retrieved textual evidence will be fed into the embedded LLM module with the input text to find the external claim-evidence inconsistencies. 
Similar to ensemble learning, the LLM module will output the final judgment and explanation, drawing on the results from both internal and external checks.
We first introduce the base MLLM and the two-stage instruction tuning procedures, before describing the reasoning process which includes internal/external checking, and composed reasoning.

\subsection{Base MLLM}
We employ InstructBLIP \cite{instructblip}, a general-purpose vision-language model as the base MLLM for further tuning. InstructBLIP consists of an image encoder, an LLM, and a Query Transformer (Q-Former). As shown in Figure~\ref{fig:sniffer},  Q-Former extracts instruction-aware visual features based on the output of the frozen image encoder, and feeds the visual features as soft prompt input to the frozen LLM.
The learnable query vectors interact with the instruction text through self-attention layers, and interact with the frozen image embeddings through cross-attention layers.
 Similar to  BLIP-2 \cite{blip2}, 
 Q-Former is pre-trained for vision-language representation learning and vision-to-language generative learning and then is tuned for multi-task vision-language instruction learning in \cite{instructblip}.

\subsection{Instruction Tuning}
\label{sec:instruction}
We use a two-stage training procedure to adapt the general-domain InstructBLIP to the news domain and OOC detection task sequentially (see  Figure~\ref{fig:training}).

\smallskip
{\noindent \bf Stage 1: News Domain Alignment.}
We observe that InstructBLIP tends to respond with coarse-grained nouns (\eg ``person'', ``woman'', and ``man'') rather than fine-grained, specific names such as ``Donald Trump''. 
OOC samples are typically created by replacing original entities with ones that are similar but not identical, and these subtle distinctions appear to be difficult for InstructBLIP to capture.

Given the cross-domain differences in the lexical preference, we construct the news-domain instruction dataset to adapt the general-domain InstructBLIP to the news domain. 
This instruction dataset is curated from the NewsCLIPpings dataset\footnote{We exclude the validation and testing set to prevent data leakage.} \cite{newsclippings} which consists of 36,8013 unique news image-caption pairs covering a diverse and representative set of news-domain concepts.
Inspired by LLaVA \cite{llava}, we leverage ChatGPT-4 \cite{chatgpt4} to construct the instruction-following data from these image-caption pairs.
Specifically, to keep the diversity of instruction data, we prompt ChatGPT-4 to generate 11 questions (see Supplementary) with the intent to instruct the model to describe the image content. 
Given that these captions have less than 30 words, we stipulate that the generated questions should explicitly specify a ``brief''
description of the images.
For an image $\mathbf{I}$ and its associated caption $\mathbf{T}_c$, we randomly sample one question $\mathbf{T}_q$ to form the corresponding instruction as 
\begin{equation}
    \text{\textbf{Human}: } \mathbf{I} \mathbf{T}_{\mathrm{q}} \langle \operatorname{STOP} \rangle; ~~ \text{\textbf{Model}:} \mathbf{T}_{\mathrm{c}}\langle \operatorname{STOP} \rangle .
\end{equation}

During training, we keep both the image encoder and LLM frozen, and only update the Q-former.
In this way, the image features of vast news-domain visual concepts can be aligned to the textual embeddings of their corresponding fine-grained entity names in the pre-trained LLM. 

\smallskip
{\noindent \bf Stage 2: Task-Specific Tuning.}
In out-of-context misinformation, an image is repurposed with a false text, which is different from the typical visual-language tasks that general-purpose MLLMs focus on. These tasks, such as image captioning and visual question answering, typically involve text and images that correspond to the same event. To address this, we construct the instruction data to progressively adapt the news-domain MLLM (developed in Stage 1) to the task of OOC misinformation detection. Similar to Stage 1, we only update the parameters in Q-former. 

\begin{figure}[t]
	\centering
	\includegraphics[width=.45\textwidth]{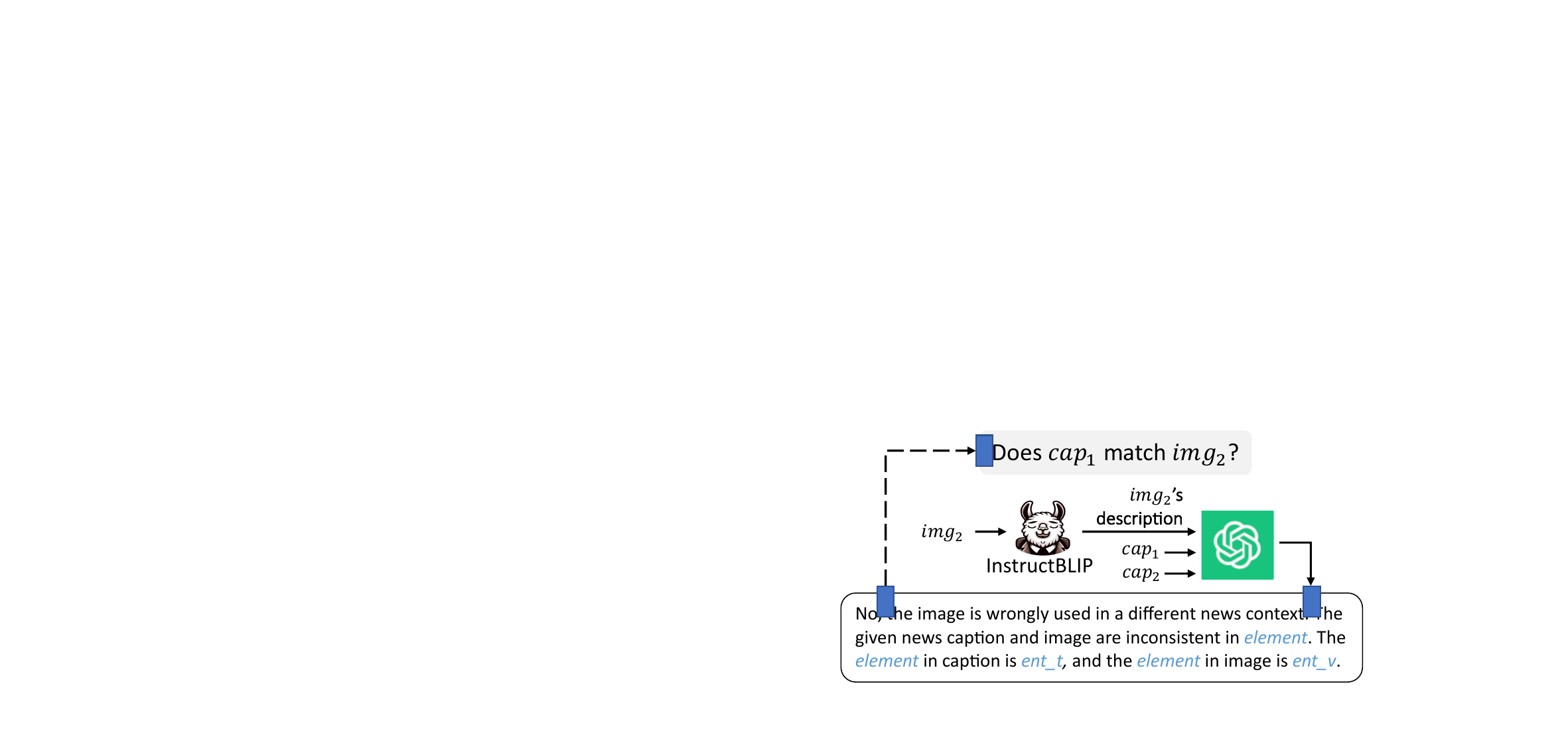}
  \vspace{-.1in}
	\caption{Process of OOC instruction generation.}
	\label{fig:oocdata}
\vspace{-.15in}
\end{figure}

One of the primary challenges in creating an explainable OOC misinformation detector is the lack of supervised data that includes both judgments and explanations.
The NewsCLIPpings dataset
generates fake pairs by replacing the $img_1$ in a real pair ($cap_1, img_1$) with another $img_2$ from a similar but different pair ($cap_2, img_2$), resulting in a fabricated pair ($cap_1, img_2$). While labels indicating their authenticity are provided for these pairs, the specific inconsistencies between $cap_1$ and $img_2$—the crux of the misinformation—are not explicitly identified.
As such, we innovatively extract the inconsistencies between $cap_1$ and $cap_2$ by prompting language-only GPT-4 as an alternative. 
As shown in Figure~\ref{fig:oocdata},
given $cap_1$, $cap_2$, and the InstructBLIP-generated  $img_2$'s basic description, we 
prompts GPT-4 to generate the inconsistencies between $cap_1$ and $img_2$ as if it could see the image (even though it only has access to the text). We curate few-shot examples in the prompt and also restrict the output format, requiring it to spot the news elements that are inconsistent between $cap_1$ and $img_2$, as well as the specific entities.
The complete prompt is shown in Supplementary. 
Although there may be multiple inconsistent elements between $cap_1$ and $img_2$, we require GPT-4 to generate only the most likely one for clarity. 
In total, we obtain 35,536 GPT-assisted instructions for out-of-context samples and supplement the real samples with an equal number of instructions (\ie, ``No, the image is rightly used in the given news context. '')

\subsection{Reasoning Process}
To effectively tackle the challenge of OOC misinformation detection, the reasoning process in \textsc{Sniffer} employs a comprehensive strategy that integrates both internal and external verification methods.

\smallskip
{\noindent \bf Internal Checking.}
The two-stage instruction tuning process equips  \textsc{Sniffer} with the ability to identify image-text inconsistency for {internal checking}. However, the model does not have access to current information as it is limited by the training corpus. Therefore, we employ Google Entity Detection API \cite{googlevision} to recognize visual entities in images which are incorporated into the instruction as supplementary information. This allows \textsc{Sniffer} to maintain relevancy and accuracy in its analysis. Additionally, there is potential to further enhance \textsc{Sniffer}'s detection capabilities by integrating it with other plug-in tools, expanding its functionality in detecting inconsistencies.

\smallskip
{\noindent \bf External Checking.}
In addition to image-text internal checking, leveraging retrieved web evidence for external verification is a crucial step, as highlighted in \cite{cheapfake-emnlp, ccn}. The context of an image, particularly the news text in which it has previously appeared, serves as a vital supplement to the content of the image itself. 
Reverse image searches can often reveal the original event where the image was first reported, to help verify if the accompanying text matches the true context of the image.
Given the strong analytical reasoning ability of LLM, we input both the news caption and the text from webpages retrieved via image search \cite{googlevision} into the LLM module in \textsc{Sniffer}. The LLM is then tasked with determining whether the provided news caption is supported by the retrieved evidence.

\smallskip
{\noindent \bf Composed Reasoning.}
The internal and external verification steps analyze the input image-text pairs from different perspectives and may yield different conclusions. 
Therefore, we employ the LLM once again for an interpretable model ensemble, tasking it to deliver its final decision based on both outcomes and the initial caption, and to clearly explain its decision-making process.

\section{Performance Study}
\label{sec:experiments}
In this section, we conducted experiments to evaluate the effectiveness of \textsc{Sniffer}. Specifically, we aim to answer the following evaluation questions:

{\noindent \bf Q1:} Can \textsc{Sniffer} improve the performance of out-of-context misinformation detection?

{\noindent \bf Q2:} How effective are the different modules of \textsc{Sniffer} in detection?

  {\noindent \bf Q3:} Can \textsc{Sniffer} generate accurate and convincing explanations for their judgment?

    {\noindent \bf Q4:} How does \textsc{Sniffer} perform in early detection, when the number of training samples is limited?
  
    {\noindent \bf Q5:} How does \textsc{Sniffer} perform on other datasets?
 
    {\noindent \bf Q6:} How does \textsc{Sniffer} perform  compared to GPT-4V?

\subsection{Experimental Setup}

{\bf Dataset.}
We use the largest out-of-context misinformation detection benchmark NewsCLIPpings \cite{newsclippings}. This dataset is built based on VisualNews \cite{visualnews}, a large-scale corpus that contains image-caption pairs from four news agencies (The Guardian, BBC, USA Today, and The Washington Post). The out-of-context samples are automatically generated by replacing the images in the original image-caption pairs with retrieved images that are semantically related but belong to different news events. \cite{ccn} extends the NewsCLIPpings dataset by supplementing the retrieved textual and visual evidence. Here, we use part of the textual evidence and detected visual entities provided in this datatset. 

Following previous works \cite{ccn, dttrm, neu-sym}, we report the results on the Merged/Balance subset, which has a balanced proportion of different retrieval strategies and positive/negative samples. The number of samples in the training, validation, and testing sets are 71072, 7024, and 7264, respectively.
As in \cite{newsclippings}, we use the accuracy over all samples (All) and separately for the Fake (Out-of-Context) and Real (Not Out-of-Context) samples as evaluation metrics.

\smallskip
{\noindent\bf Implementation Details.}
We select InstructBLIP \cite{instructblip} as the base MLLM, of which the image encoder is ViT-G/14 \cite{vit} and the LLM is Vicuna-13B \cite{vicuna}. During training, we initialize the model from the pre-trained InstructBLIP and only finetune the parameters of Q-Former while keeping both the image encoder and the LLM frozen. To reduce the memory cost, we use FlashAttention-2 \cite{flashatt} to replace the standard Attention layer in LLM. 
Our implementation uses the LAVIS library \cite{lavis}. The batch size is set as 8 and 4 in the stages of news domain alignment and task-specific tuning, respectively. The max input sequence length is 550 and the output length is 256. 
We use the AdamW \cite{adamw} optimizer and apply a linear warmup of the learning rate, increasing from $10^{-8}$ to $10^{-5}$, followed by a cosine decay.
The models are trained utilizing 4 Nvidia A100 (40G) GPUs.

\smallskip
{\noindent\bf Baselines.}
We compared \textsc{Sniffer} with two representative multimodal misinformation detectors trained from scratch: 1) {\bf SAFE} \cite{safe} which translates the input image into a sentence, and computes the multimodal relevance based on the sentence similarity as the auxiliary loss; 2) {\bf EANN} \cite{eann} which uses adversarial training to guide the model to learn event-invariant multimodal features for detection.

We also compared \textsc{Sniffer} with  pre-trained multimodal baselines:
1) {\bf VisualBERT} \cite{visualbert}, one of the earliest works on multimodal pre-training, concatenates the bounding box features and textual tokens. The combined features are fed into a unified encoder, consisting of a series of transformer layers, to align them into one embedding space. The pre-training of VisualBERT involves  masked language modeling and image-text matching objectives;
2) {\bf CLIP} \cite{clip} passes image and text through separate encoders and uses contrastive loss to guide the multimodal encoders to generate similar representations for related concepts;
3) {\bf DT-Transformer} \cite{dttrm} uses CLIP as the multimodal encoder and adds auxiliary Transformer layers to enhance the multimodal features interaction. It combines different types of generated data as the training corpus;
4) {\bf CCN} \cite{ccn} proposes a consistency-checking network assisted by CLIP, that considers the consistency of the claim-evidence (image-image and text-text), in addition to the image-text pairing. This is the only baseline that utilizes retrieved external evidence;
5) {\bf Neu-Sym detector} \cite{neu-sym} proposes a neural-symbolic model which symbolically disassembles the text into a set of fact queries, and forwards the query-image pairs into a pre-trained MLLM. The output answers are further selected and combined to obtain the final judgment.

\subsection{Performance Comparison (Q1)}
 Table~\ref{tab:comparison} shows the performance of \textsc{Sniffer} and the various baselines. We observe that
\textsc{Sniffer} outperforms all the baselines, which validates that it can effectively capture the subtle inconsistency in the OOC samples.  
Even though \textsc{Sniffer} only considers part of the textual evidence, it still surpasses CCN by over 3.7\%.
Baselines trained from scratch (\ie EANN and SAFE) are worse than pre-trained multimodal baselines, verifying the importance of world knowledge in the pre-trained models in distinguishing the OOC samples. 
The inferior performance of the Neu-Sym detector compared to other approaches (\eg DT-Transformer) suggests that general-purpose MLLM may not be well-suited for the OOC detection task.

\begin{table}[t]
\small
\centering
\caption{Performance (\%) comparison between \textsc{Sniffer} and baselines. The best results are in {\bf boldface}. }
\vspace{-.1in}
\begin{tabular}{lccc}
\toprule
\makecell[c]{\textbf{Method}}  & \textbf{All} & \textbf{Fake} & \textbf{Real} \\
\midrule
SAFE & 52.8 & 54.8 & 52.0 \\ 
EANN & 58.1 & 61.8 & 56.2 \\ 
VisualBERT & 58.6 & 38.9 & 78.4 \\ 
CLIP & 66.0 & 64.3 & 67.7 \\ 
DT-Transformer & 77.1 & 78.6 & 75.6 \\
CCN & 84.7 & 84.8 & 84.5 \\
Neu-Sym detector & 68.2 & -  & - \\ 
\textsc{Sniffer} ({\it Ours}) & {\bf 88.4} & {\bf 86.9} & {\bf 91.8} \\         
\bottomrule
\end{tabular}
\label{tab:comparison}
\end{table}

\subsection{Ablation Studies (Q2)}
We conducted ablation experiments to analyze the importance of each component of \textsc{Sniffer} in detecting OOC misinformation. 
Specifically, we first tested the original {\bf InstructBLIP}, and then incrementally integrated various components: pre-training ({\bf PT}) with news-domain data in Stage 1, task-specific tuning based on OOC data in Stage 2 ({\bf OOC tuning}), visual entities ({\bf VisEnt}), and retrieved external evidence ({\bf Evidence}). 
From Table~\ref{tab:ablation}, we can see that:

\begin{table}[t]
\begin{center}
\caption{Ablation studies on each component in \textsc{Sniffer}. }
\vspace{-.1in}
\label{tab:ablation}
\footnotesize
\setlength{\tabcolsep}{3pt}
\begin{tabular}{ccccc|ccc}
\toprule
InstructBLIP & PT & OOC Tuning & VisEnt & Evidence & {\bf All} & {\bf Fake} & {\bf Real} \\
\midrule
\textcolor{dark-green}{\cmark} & \xmark & \xmark & \xmark & \xmark & 47.4 & 4.6 & 90.3 \\
\textcolor{dark-green}{\cmark} & \textcolor{dark-green}{\cmark} & \xmark & \xmark & \xmark  & 49.3 & 9.4 & 89.2 \\
\textcolor{dark-green}{\cmark} & \xmark & \textcolor{dark-green}{\cmark} & \xmark & \xmark & 82.5 & 75.3 & 89.7 \\
\textcolor{dark-green}{\cmark} & \xmark & \textcolor{dark-green}{\cmark} & \textcolor{dark-green}{\cmark} & \xmark & 87.6 & 83.9 & 91.3 \\
\textcolor{dark-green}{\cmark} & \textcolor{dark-green}{\cmark} & \textcolor{dark-green}{\cmark} & \xmark & \xmark & 83.1 & 76.5 & 89.6 \\
\textcolor{dark-green}{\cmark} & \textcolor{dark-green}{\cmark} & \textcolor{dark-green}{\cmark} & \textcolor{dark-green}{\cmark} & \xmark & 88.2 & 84.9 & 94.0 \\
\textcolor{dark-green}{\cmark} & \xmark & \xmark & \xmark & \textcolor{dark-green}{\cmark} & 84.5 & {\bf 92.9} & 76.0 \\
\textcolor{dark-green}{\cmark} & \textcolor{dark-green}{\cmark} & \textcolor{dark-green}{\cmark} & \textcolor{dark-green}{\cmark} & \textcolor{dark-green}{\cmark} & {\bf 88.4} & 86.9 & {\bf 91.8} \\
\bottomrule
\end{tabular}
\end{center}
\vspace{-.3in}
\end{table}

\begin{itemize}
    \item The original InstructBLIP exhibits a classification accuracy of only 47.4\%, which is even lower than random guessing. Furthermore, the recall for fake samples was a mere 4.6\%. These results indicate that the general-purpose InstructBLIP tends to misclassify OOC misinformation as real, likely due to the similarity in the text and image composition of such challenging misinformation.
    \item All components in \textsc{Sniffer} are important for achieving its best performance, especially the OOC tuning. Specifically, it improves the performance of InstructBLIP by over 35 percentage points, suggesting that task-specific tuning can aid the model in learning the logic necessary to accurately judge OOC samples.
    \item The incorporation of visual entities contributed to a 5-point increase in accuracy.
    This indicates that the integration of external tools can significantly enhance the model's ability to identify visual elements.
    \item Relying solely on assessing the relevance between text and image-retrieved textual evidence can yield relatively good detection results. However, while this approach achieves high recall for fake samples, that of real samples is notably lower than other methods. This discrepancy is largely attributable to the noise in the retrieved evidence. Even for real news, relevant evidence may not always be retrievable, leading to misclassification of real samples. Additionally, since only less than 60\% of the samples have associated evidence, the overall improvement in the model's accuracy is somewhat limited.
\end{itemize}

\subsection{Explainability Analysis (Q3)}
We evaluated the quality of the explanations generated by \textsc{Sniffer} in terms of accuracy and persuasiveness.

{\noindent \bf Quantitative Analysis.}
We compared \textsc{Sniffer}'s generated explanation with the ground truth obtained in Section \ref{sec:instruction} on the test set.
We focus on three critical information points within OOC samples, \ie inconsistent news {\it element}, the entity of that element in the text ({\it ent\_t}) and image ({\it ent\_v}). 
We design two types of evaluation metrics: 1) Response ratio of these three points; 2) Accuracy of these three points. We use hard match and compute the hit ratio for {\it element}, and use similarity of the CLIP embedding for {\it ent\_t} and {\it ent\_v}. 
We also use ROUGE \cite{rouge}  to measure the accuracy of the generated explanations from an overall perspective.  
We compared the performance of various variants of \textsc{Sniffer} on these metrics to demonstrate the role of different components in enhancing the model's explainability.

\begin{figure}[t!]
	\centering
		\centering
		\includegraphics[width=0.8\columnwidth]{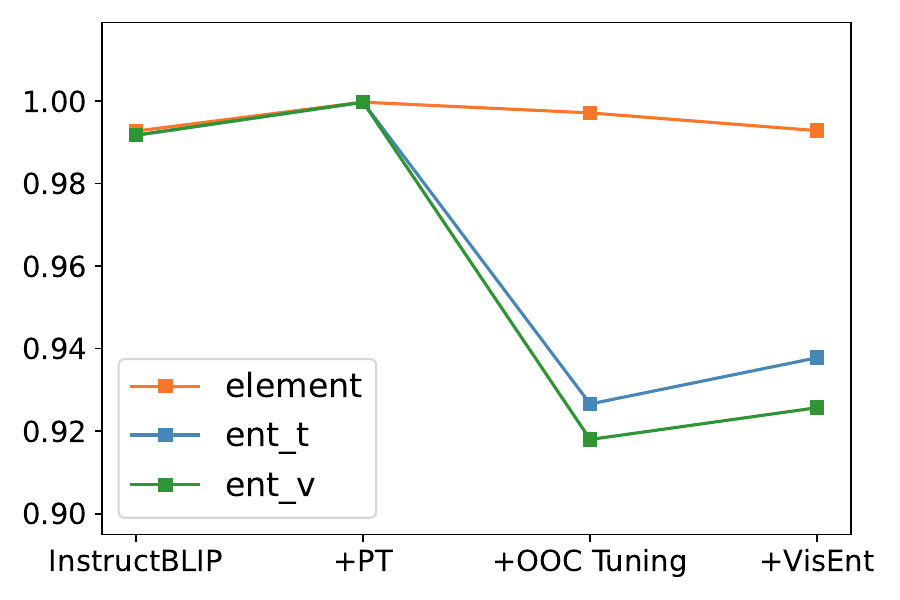}
  \vspace{-.11in}
  \caption{Response ratio.}
    \vspace{-.15in}
		\label{fig:response}
	\end{figure}
	\begin{figure}[t!]
		\centering
		\includegraphics[width=0.8\columnwidth]{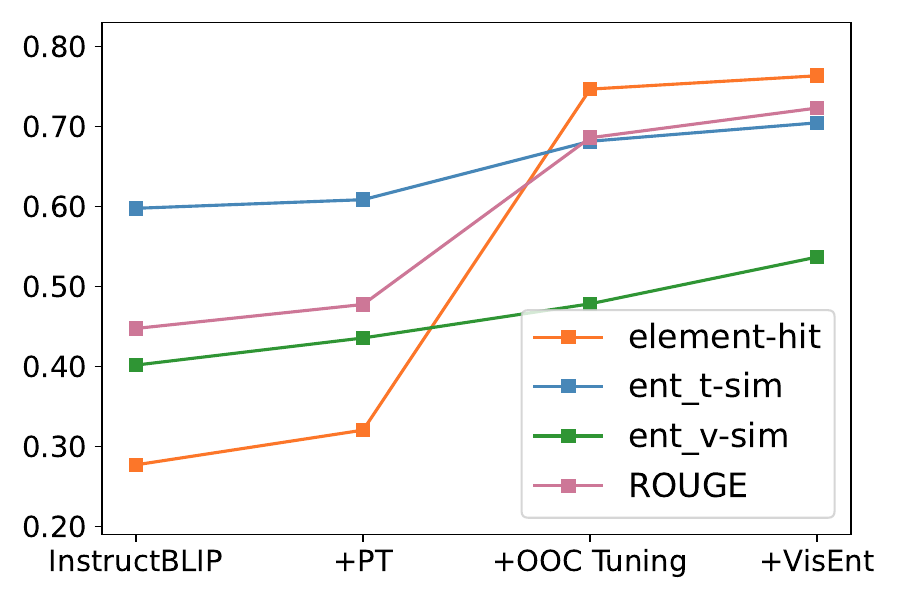}
  \vspace{-.11in}
		\caption{Explanation accuracy. }
		\label{fig:acc}	
    \vspace{-.15in}
\end{figure}

{\bf 1) Response Ratio:} Figure~\ref{fig:response} shows the response ratio of {\it element}, {\it ent\_t}, and {\it ent\_v} for different training steps. We see that the model's response ratio is close to 1 initially but significantly decreases for {\it ent\_t} and {\it ent\_v} after OOC tuning. This indicates that the model has become more conservative in its outputs. Further, the response ratio improves slightly after incorporating visual entities, indicating that these entities have enhanced the model's ability to recognize entities.

{\bf 2) Explanation Accuracy: } Although there is a decline in the response ratio of the model, the accuracy of its responses across all three points shows an increase. Figure~\ref{fig:acc} shows that all components contribute to the improvement of explanation accuracy in terms of the hit ratio for {\it element}, average similarity for {\it ent\_t} and {\it ent\_v}, and ROUGE value for the whole response.
Specifically, the hit ratio of {\it element} increases by 4\% and 44\% by pre-training and OOC tuning with visual entities, respectively, demonstrating the model's increasing ability to capture inconsistencies between text and images. 
The accuracy of {\it ent\_t} consistently outperforms {\it ent\_v} by 17\%. This disparity can be attributed to the relative simplicity of directly extracting {\it ent\_t} from text, as opposed to {\it ent\_v}, which is derived from images and thus heavily dependent on the model's visual recognition capabilities.

Overall, after training, the model has become more conservative yet more accurate in spotting the key detection points. This indicates that the model has truly captured clues of inconsistencies between text and images, rather than merely fitting to real or fake labels.

\begin{figure}[t]
	\centering
	\includegraphics[width=.43\textwidth]{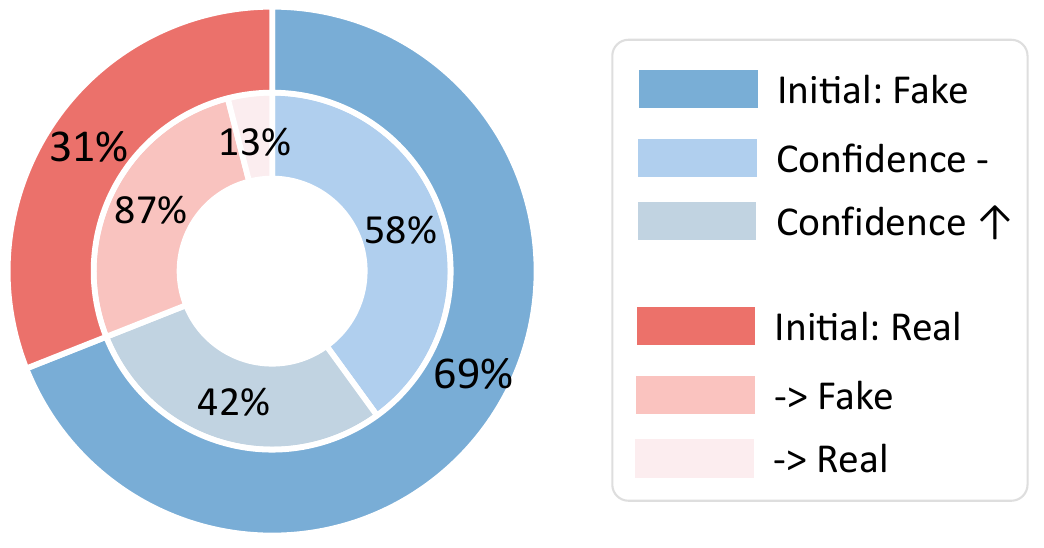}
 \vspace{-.07in}
	\caption{Results of human evaluation. The outer ring displays the initial distribution of test samples judged as real and fake, while the inner circle illustrates the changes in users' judgments and confidence levels after reading \textsc{Sniffer}'s explanations. }
	\label{fig:user}
  \vspace{-.1in}
\end{figure}

\smallskip
{\noindent\bf Human Evaluation.} 
To assess the effectiveness of \textsc{Sniffer} in debunking misinformation through its generated explanations, we conducted a human evaluation. We randomly selected 20 OOC misinformation samples from the test set. Ten participants were asked to judge the veracity of each news item (\ie fake or real) and their confidence level (\ie no, somewhat, high) both before and after reading \textsc{Sniffer}'s explanations. Figure~\ref{fig:user} shows that: (a) 69\% of the items were correctly identified as fake by the users\footnote{When engaged in data annotation, users tended to label news items as fake more frequently compared to their usual behavior when reading news on social media.}, a finding that aligns with the statistics in \cite{newsclippings}, indicating a certain level of user discernment in detecting OOC misinformation; (b) for the OOC samples initially misidentified as real by the users, 87\% of their judgments changed to fake after reading \textsc{Sniffer}'s judgment and explanations, demonstrating the strong persuasive power of \textsc{Sniffer}; (c) for the OOC samples that were initially identified as fake by the users, \textsc{Sniffer}'s explanations also increased the users' confidence in their judgments for 42\% of these samples.

\begin{figure}[t!]
\centering
\includegraphics[width=0.8\columnwidth]{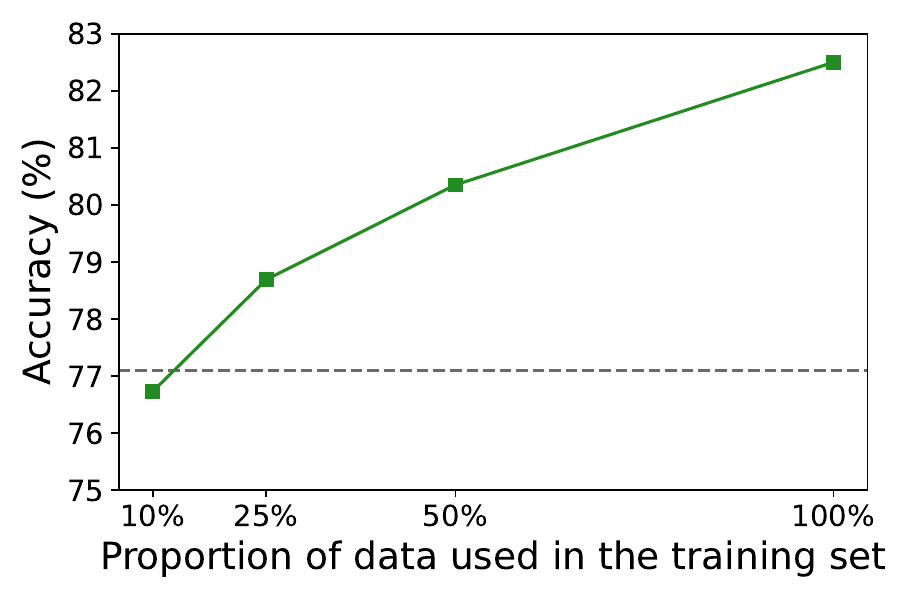}
  \vspace{-.12in}
		\caption{Performance in early detection.}
		\label{fig:ratio}
    \vspace{-.15in}
\end{figure}

\subsection{Practical Setting}
\noindent {\bf Early Detection (Q4). }
Detecting misinformation in its early stage is important for timely mitigating its negative influences \cite{early}. We conducted experiments using different proportions of the training set to evaluate the model's performance. Specifically, we randomly selected 10\%, 25\%, 50\% and 100\% training samples and conducted experiments on \textsc{Sniffer-}, a reduced version of \textsc{Sniffer} that only employs OOC Tuning on InstructBLIP, for a fair comparison.  
Figure~\ref{fig:ratio} shows that, with merely 10\% of the training data,
\textsc{Sniffer-}  achieves comparable detection performance as baseline models that do not utilize external evidence.
When trained with 25\% of the data, it 
can surpass their performance,  underscoring the superiority of MLLM in terms of training efficiency and detection accuracy.

\begin{figure}
     \centering
     \begin{subfigure}[b]{0.235\textwidth}
         \centering
         \includegraphics[width=\textwidth]{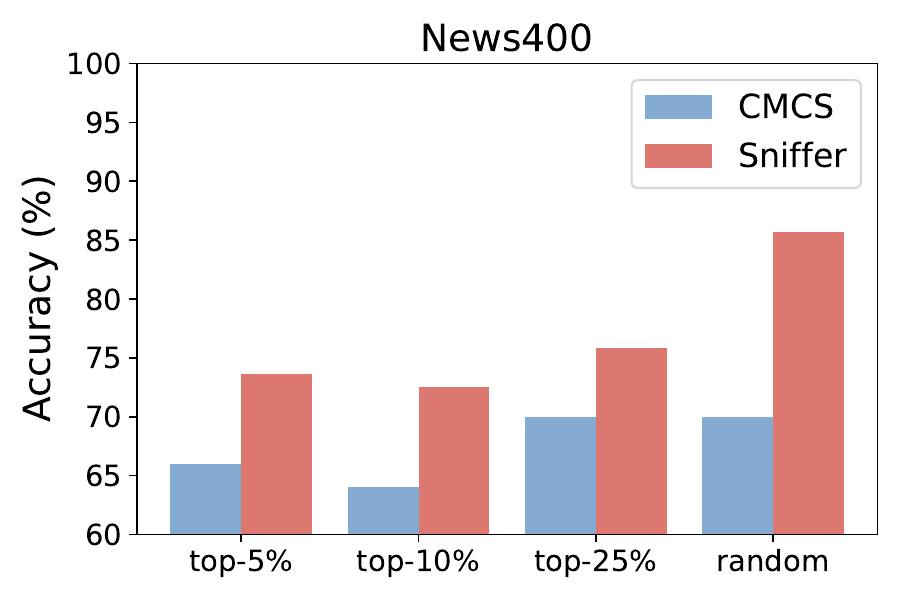}
         \caption{}
           \vspace{-.1in}
         \label{fig:news400}
     \end{subfigure}  
     \begin{subfigure}[b]{0.235\textwidth}
         \centering
         \includegraphics[width=\textwidth]{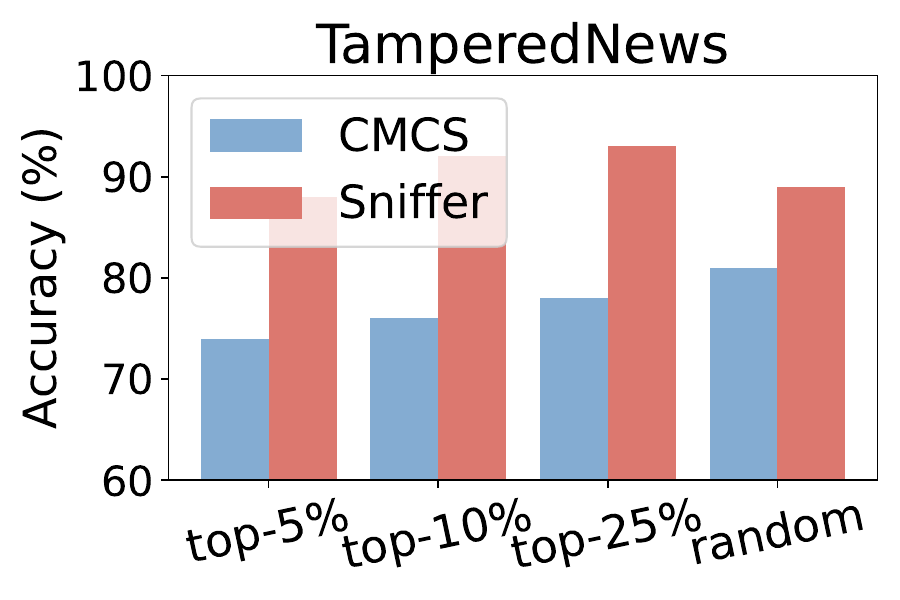}
         \caption{ }
           \vspace{-.1in}
         \label{fig:tampered}
     \end{subfigure}
        \caption{Cross-dataset detection performance of \textsc{Sniffer}. }
        \label{fig:cross-dataset}
          \vspace{-.15in}
\end{figure}

\smallskip
\noindent {\bf Generalization Analysis (Q5)}.
To validate the generalizability of \textsc{Sniffer}, we tested the model that was trained on the NewsCLIPpings dataset, on the other two datasets, \ie News400 and TamperedNews datasets \cite{ijmir}. 
$img_1$ is replaced by images of similar appearance ($img_2$) based on top-$k$ similarity to $img_1$, creating different subsets according to the value of $k$. A smaller $k$ indicates higher similarity between $img_1$ and $img_2$, and hence, a greater challenge in detection.
We compared \textsc{Sniffer} with 
Cross-Modal Context Similarity (CMCS), a competitive baseline proposed in \cite{ijmir}. CMCS quantifies the similarity between the textual context and visual scene labels within the textual feature space. 
Figure~\ref{fig:cross-dataset} shows that \textsc{Sniffer} outperforms CMCS by a large margin at different difficulty levels in both datasets,  confirming the cross-dataset generalizability of \textsc{Sniffer}.

\subsection{Comparison with GPT-4V (Q6)}
As the most popular MLLM, GPT-4 with vision (GPT-4V) \cite{gpt4v} has demonstrated unparalleled performance across a variety of multimodal tasks, which inspired us to test it on the OOC detection task. We randomly sampled 400 samples from the test set (200 fake and 200 real examples) and queried GPT-4V  using the same prompts as those used for \textsc{Sniffer}. Table~\ref{tab: gpt} shows that \textsc{Sniffer} outperforms GPT-4V by 11\% in terms of classification accuracy. 
This demonstrates that task-specific (relatively) smaller models are fully capable of outperforming general-purpose larger models in specific tasks\footnote{Currently, there is no literature explicitly stating the size of GPT-4V, but it is anticipated to exceed the 175 billion parameter scale of GPT-3 \cite{gpt3}.}.
To further understand the behavioral differences between \textsc{Sniffer} and GPT-4V, we compare their generated explanations in Supplementary.

\begin{table}[h]
\centering
\small
\caption{Comparison of classification accuracy (\%) between \textsc{Sniffer} and GPT-4V on randomly sampled test set. }
 \vspace{-.05in}
\begin{tabular}{lccc}
\toprule
\makecell[c]{\textbf{Method}}  & \textbf{All} & \textbf{Fake} & \textbf{Real} \\
\midrule
GPT-4V & 75.5 & 77.0 & 74.0 \\ 
\textsc{Sniffer} ({\it Ours}) & {\bf 86.8} & {\bf 79.0} & {\bf 94.5} \\         
\bottomrule
\end{tabular}
\label{tab: gpt}
 \vspace{-.15in}
\end{table}
\section{Conclusion }
\label{sec:conclusion}
In this paper, we proposed \textsc{Sniffer}, a multimodal large language model for out-of-context misinformation detection, providing both judgment and explanation. To develop \textsc{Sniffer}, we constructed multi-perspective instruction data assisted by GPT-4, and employed instruction tuning to continuously adapt the general-purpose InstructBLIP for the news domain and OOC detection task. We further augmented the model by integrating external tools and retrieval methods. Experiments prove that \textsc{Sniffer} not only achieves SOTA performance in detection, but also generates precise and persuasive explanations.

\appendix
\renewcommand\thesection{\Alph{section}}
\lstset{escapeinside=`}

\renewcommand\arraystretch{1.5}
\begin{table*}[htbp]
\small
\caption{Behaviors of existing open-sourced general-purpose MLLMs in detecting OOC misinformation. }
\vspace{-.1in}
\begin{tabular}{p{2cm}m{14cm}}
\toprule
User & Some rumormongers use images from other events as illustrations of the current news event to make up
multimodal misinformation. Given a news caption and a news image, judge whether the given
image is rightly used in the given news context. Let’s analyze their inconsistency from
perspectives of main news elements, including time, place, person, event, artwork, etc. You
should answer in the following forms: ``Yes, the image is rightly used." or ``No, the image is
wrongly used in a different news context. The given news caption and image are inconsistent
in \textless element\textgreater. The \textless element\textgreater in the caption is \textless ent\_t\textgreater, and the \textless element\textgreater in the image is \textless ent\_v\textgreater."
News caption: Saturn V booster was used in Nasa space missions between 1967 and 1972. 
The answer is \\
& \makecell[c]{\includegraphics[width=.3\textwidth]{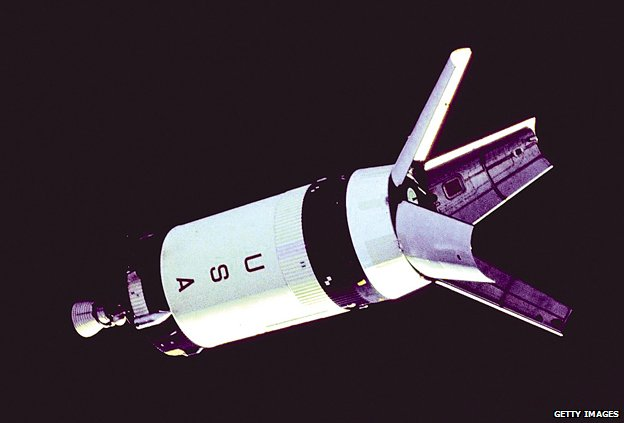}} \\ 
InstructBLIP & 1967-1972, which is the time period when the Saturn V booster was used in Nasa space missions.  \\ 
LLaVA-Adapter-V2 & No, the image is wrongly used in a different news context. The image is of a Saturn V rocket, which was used in NASA space missions between 1967 and 1972. However, the given news caption refers to a different event, the Apollo 11 mission, which took place in 1969. The image and caption are inconsistent in time, as the Saturn V rocket was not used in the Apollo 11 mission. \\ 
Woodpecker & No, the image is wrongly used in a different news context. The given news caption and image are inconsistent in time. The caption states that the Saturn V booster was used in Nasa space missions between 1967 and 1972, while the image depicts a rocket launching in the 1980s. The image is not related to the Saturn V booster, which was used in the 1960s and 1970s.  \\ 

\bottomrule
\end{tabular}
\label{tab:behaviors}
\end{table*}

\begin{table*}[t]
\centering
\caption{The list of instructions for brief news image description.}
\vspace{-.2in}
\begin{tcolorbox}[width=.65\textwidth]
\centering
\small
\begin{itemize}[leftmargin=.1mm]
\setlength{\itemsep}{2pt}
\item ``Analyze the news image and provide a brief summary of the event it depicts.''
\item ``Give a short description of the scene shown in the news photograph.''
\item ``Offer a concise report based on the news-related image provided.''
\item ``Summarize the news story as represented by the image.''
\item ``Interpret the journalistic image and detail the key elements in a succinct manner.''
\item ``Provide a brief journalistic overview of the news picture.''
\item ``Construct a short narrative to convey the news event shown in the image.''
\item ``Deliver a compact exposition of the incident captured in the news photo.''
\item ``Express the main news points illustrated by the given photograph.''
\item ``Condense the news context of the image into a clear, brief description.''
\item ``What news event does this image describe?''
\end{itemize}
\end{tcolorbox}
    
\label{tab:11question}
\end{table*}

\begin{figure*}[t]
\lstset{style=mystyle,
        frame=none,
        keywordstyle = \color{black}, 
        commentstyle =\color{codegreen}, 
        stringstyle = \color{black}, 
        breakindent=0\textwidth,
        frame = single,
        backgroundcolor=\color{white},
        xleftmargin=0.05\textwidth,
        xrightmargin=0.05\textwidth}
\begin{lstlisting}[language=Python]
# system message
You are an expert in fact-checking. Some news captions and accompanying images are inconsistent in terms of key news elements (5W1H) because rumormongers have taken images from other news and used them as illustrations for current news to make up multimodal misinformation. Given the original news caption (i.e. caption_ori) for the image (i.e. image_ori), the new news caption (i.e. caption_new), and a basic description of image_ori's content, I need you to analyze the inconsistencies between caption_ori and caption_new in key news elements, and select from them the one most likely inconsistency between image_ori and caption_new based on the description of image_ori. You should answer in the following forms:  "They are inconsistent in <element>. The <element> in caption_new is <ent_t>, and the <element> in image_ori is <ent_v>.\n Element: <element> \n Entity_caption: <ent_t> \n Entity_image: <ent_v>\". Please answer only one inconsistent element.

(*@\color{codegreen}{\# in-context examples}@*)
Caption_ori:  John Constable's Brightwell Church and Village was part of the 2013 exhibition
Caption_new:  From J Charles Eichhorn's American Skat or The Game of Skat Defined
Basic description of image_ori: This image describes a rural landscape with a farmhouse, a barn, and a field. The farmhouse is situated in the middle of the field, surrounded by the barn and the open land. The painting captures the essence of a peaceful, pastoral setting, with the farmhouse serving as the central focus of the scene.
The answer is: They are inconsistent in artwork. The artwork in caption_new is American Skat or The Game of Skat Defined, and the artwork in image_ori is Brightwell Church and Village. \n Element: artwork \n Entity_caption: American Skat or The Game of Skat Defined \n Entity_image: Brightwell Church and Village

Caption_ori:  Chris Huhne is among the ministers expected to address delegates at next week's Lib Dem conference
Caption_new:  Urs Rohner CEO of Credit Suisse participates in a panel session in Bern Switzerland on Tuesday
Basic description of image_ori: The image depicts a man wearing a suit and tie, standing at a podium with a microphone in front of him. He appears to be giving a speech or addressing an audience. In the background, there is a black screen or backdrop.
The answer is: They are inconsistent in person. The person in caption_new is Urs Rohner, and the person in image_ori is Chris Huhne. \n Element: person \n Entity_caption: Urs Rohner \n Entity_image: Chris Huhne

# query
Caption_ori:  (*@{\color{codepurple}{\textsc{<caption\_ori>}}@*)  
Caption_new:  (*@{\color{codepurple}{\textsc{<caption\_new>}}@*)  
Basic description of image_ori:  (*@{\color{codepurple}{\textsc{<basic\_description\_of\_image\_ori>}}@*) 
The answer is: 

\end{lstlisting}
\caption{Prompt used to ask GPT-4 to generate the OOC instruction data. }
\label{fig:chatgpt}
\end{figure*}

\renewcommand\arraystretch{1}
\begin{table*}[t]
\small
\caption{Comparison between GPT-4V and \textsc{Sniffer}. For each case, we annotate the ground truth (\textcolor{lightblue}{GT}) and the correctness of the model's judgment, indicated by either \textcolor{codegreen}{\checkmark}(correct) or \textcolor{red}{\xmark} (wrong).}
\begin{tabular}{lll}
\toprule
\multicolumn{2}{p{12cm}}{{\bf Caption}: Nick Clegg (left) in Oldham with Liberal Democrat candidate Elwyn Watkins (centre) and Lib Dem deputy leader Simon Hughes. \color{lightblue}{[GT: FAKE]}} & \makecell[c]{\multirowcell{14}{\includegraphics[width=.25\textwidth]{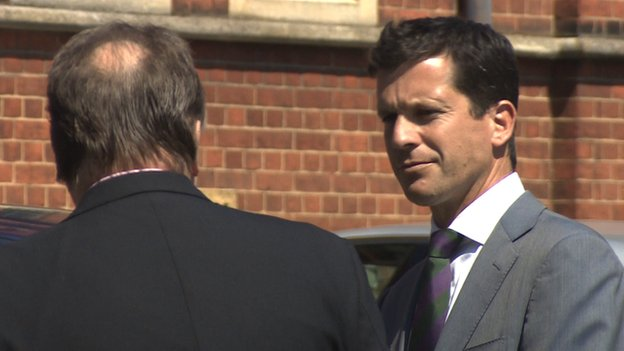}}}  \\
\cmidrule{1-2}
\multicolumn{1}{p{6cm}|}{{\bf GPT-4V}:  Yes, the image is rightly used. \color{red}{\xmark}}  
&  
\multicolumn{1}{p{6cm}}{{\bf \textsc{Sniffer}}: No, the image is wrongly used in a different news context. On the one hand, the person in the caption,  Nick Clegg, Simon Hughes and Elwyn Watkins, do not match the person in the image, Tim Henman. On the other hand, the caption refers to Nick Clegg, Simon Hughes, and Elwyn Watkins attending an event, while the image-retrieved webpages are about the funeral of a tennis player named Elena Baltacha. The two entities are not connected, and there are no common elements or context that link them. Therefore, the image is more likely to be wrongly used in the caption. \color{codegreen}{\checkmark}} \\
\midrule
\midrule
\multicolumn{2}{p{12cm}}{{\bf Caption}: People cheer and take pictures as the pope arrives to meet with representatives of the World of Work organization in Ciudad Juarez. \color{lightblue}{[GT: FAKE]}} & \makecell[c]{\multirowcell{15}{\includegraphics[width=.25\textwidth]{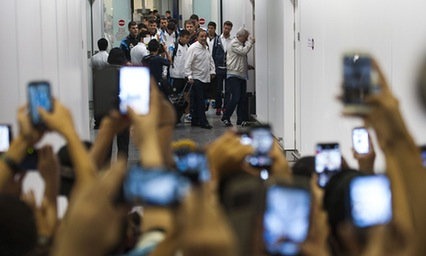}}}  \\
\cmidrule{1-2}
\multicolumn{1}{p{6cm}|}{{\bf GPT-4V}:  No, the image is wrongly used in a different news context. The given news caption and image are inconsistent in person. The person in the caption is the pope, and the person in the image is a man who does not appear to be the pope. The man seen in the image is not dressed in papal attire, which is distinctive and would typically include white robes and sometimes a zucchetto (skullcap) for the pope. Additionally, the context of the picture does not seem to match the typical scenarios where the pope is present, which often feature heightened security and more ceremonial settings, depending on the occasion. \color{codegreen}{\checkmark}}  &  
\multicolumn{1}{p{6cm}}{{\bf \textsc{Sniffer}}: Yes, the image is rightly used.  \color{red}{\xmark}} \\
\midrule\midrule
\multicolumn{2}{p{12cm}}{{\bf Caption}: Passengers are stranded outside the Guangzhou railway station in China's Guangdong province on Tuesday. \color{lightblue}{[GT: REAL]}} & \makecell[c]{\multirowcell{16}{\includegraphics[width=.25\textwidth]{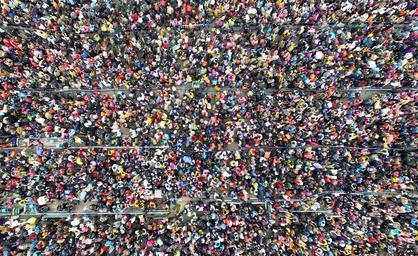}}}  \\
\cmidrule{1-2}
\multicolumn{1}{p{6cm}|}{{\bf GPT-4V}: No, the image is wrongly used in a different news context. The event in the caption is passengers stranded outside the Guangzhou railway station, and the event in the image is a large gathering of people at an event that is not related to a transportation disruption. This can be inferred from the fact that the people appear to be standing very close to one another in an organized manner, rather than having the disorganized, possibly frustrated behavior one might expect from stranded passengers. Additionally, there are no clear signs of a railway station, such as platforms, tracks, or trains, visible in the image. The setting does not match the context described in the news caption.  \color{red}{\xmark}}  
&  
\multicolumn{1}{p{6cm}}{{\bf \textsc{Sniffer}}: Yes, the image is rightly used. On the one hand, the image depicts a group of people, which is consistent with the caption. On the other hand, the image-retrieved webpages mention the Guangzhou railway station in China, the Chinese New Year and a situation where passengers were stranded outside the railway station due to snow, which are relevant to the caption. Therefore, the image is likely to be correctly used in the caption.   \color{codegreen}{\checkmark}} \\
\bottomrule
\end{tabular}
\label{tab:cases}
\end{table*}

\section{Behaviors of open-sourced general-purpose MLLMs}
\label{sec:behavior}
We evaluated three representative open-sourced general-purpose MLLMs, including InstructBLIP \cite{instructblip}, LLaVA-Adapter-V2 \cite{llava-adapter} and Woodpecker \cite{woodpecker}, for their performance in detecting out-of-context misinformation.
As demonstrated in Table~\ref{tab:behaviors}, despite explicit instructions regarding the desired output format, these MLLMs exhibit a {\it failure to follow instructions accurately}.
InstructBLIP, in particular, does not provide a clear judgment, instead merely paraphrasing the given caption, which indicates a {\it misunderstanding of the user's intent}.

Furthermore, both LLaVA-Adapter-V2 and Woodpecker display {\it hallucination} phenomena: LLaVA-Adapter-V2 erroneously attributes the content of the caption to the image and assigns a new, erroneous meaning (\ie ``the Apollo 11 mission, which took place in 1969'') to the caption; Woodpecker attributes an incorrect new context to the image (\ie ``a rocket launching in the 1980s''). This hallucination effect may be attributed to the training data, where text and image are generally aligned to depict the same event.

We also analyzed InstructBLIP's lexical preferences based on the descriptions it generated for images within the NewsCLIPpings dataset. Our statistics reveal that in samples containing person nouns, only 27\% of the responses utilize fine-grained proper nouns (\ie specific names of individuals), while the remaining 73\% employ coarse-grained common nouns (such as ``person'', ``woman'' and ``man''). 
This suggests that InstructBLIP tends to favor more general nouns over specific proper nouns in its responses.

\section{Instruction Data Construction}
\label{sec:prompt-instruction}

\paragraph{Instructions for brief image description.}
Table~\ref{tab:11question} shows the ChatGPT-generated questions to construct the diverse instruction data for news domain alignment. They present the same meaning with natural language variance.

\paragraph{Prompt to generate the OOC instruction. }
Figure~\ref{fig:chatgpt} illustrates the prompt utilized for asking GPT-4 to identify inconsistencies between $cap_1$ and $cap_2$. From the responses, we extract $element$, $ent\_t$, and $ent\_v$ to formulate the OOC instruction data, as depicted in Figure~\ref{fig:oocdata} in the main paper.

\section{Case Studies}
\label{sec:case}

We show three representative cases in Table~\ref{tab:cases} to reveal the behavioral differences between GPT-4V and \textsc{Sniffer}. 

{\it In the first example}, \textsc{Sniffer} identifies the claim as fake based on both the image-text inconsistency and the claim-evidence irrelevance. In contrast, GPT-4V fails to detect any inconsistencies. This highlights \textsc{Sniffer}'s superiority in recognizing news entities and utilizing external knowledge.

{\it In the second example}, the image depicts a scene that aligns with the caption's description of ``people cheer and take pictures'' and no relevant webpages were retrieved for this image, leading \textsc{Sniffer} to incorrectly classify this claim as real. However, GPT-4V, drawing on the Pope's attire and the security conditions at the event, deduces that the image does not depict a scene of the Pope participating in an event, and thus correctly classifies the claim as fake. This example demonstrates GPT-4V's superior world knowledge and reasoning capabilities, far surpassing those of smaller models.

{\it The third example} presents a real news story about passengers stranded outside the Guangzhou railway station, accompanied by an image of people queuing. \textsc{Sniffer}, evaluating both the image-text consistency and the claim-evidence relevance, correctly identifies this news as real. In contrast, GPT-4V erroneously classifies this news as fake, reasoning that the orderly crowd in the image does not resemble the expected chaos of stranded passengers and noting the absence of any railway station signage. This instance illustrates GPT-4V's overly cautious and sensitive judgment, predicated on the expectation that key elements of a real news story should be visibly represented in the accompanying image. 
In fact, existing research \cite{acl-mmrelation} has found that more than half of social media news stories do not have their content words represented in the images. 
This tendency also accounts for GPT-4V's lower recall rate for real news samples, as outlined in Table~\ref{tab: gpt} in the main paper.

In summary, \textsc{Sniffer} integrates clues from both text-image inconsistency and claim-evidence relevance, leading to more comprehensive judgments. Benefiting from task-specific tuning, it demonstrates a deeper understanding of the complex logic inherent in the OOC detection task. On the other hand, GPT-4V, with its vast repository of world knowledge, is adept at detecting subtle anomalies, yet this same attribute can lead to the misclassification of real news as fake.

{
    \small
    \bibliographystyle{ieeenat_fullname}
    \bibliography{main}
}

\end{document}